\documentclass[twocolumn,10pt]{article} 

\usepackage[square,numbers,sort&compress,comma]{natbib}

\usepackage{amsmath}
\usepackage{amssymb}
\usepackage{caption}
\usepackage{graphicx}
\usepackage{latexsym}
\usepackage{times}

\usepackage{color}
\usepackage{amsfonts}
\usepackage{mathrsfs}
\usepackage[separate-uncertainty]{siunitx}
\usepackage{subcaption}

\topmargin - 12pt 
\renewenvironment{abstract}%
              {
               \small
               {\bfseries \abstractname}
               \par
               \vspace{10pt}
              }

\renewcommand\abstractname{Abstract}

\newcommand{\nomenclature}
              [1]
              {
               \bgroup
               \flushleft
               \small\bf
               #1
               \par
               \egroup
              }

\renewcommand{\section}
              [1]
              {
               \bgroup
               \flushleft
               \small\bf
               \stepcounter{section}
               \arabic{section}. #1
               \par
               \egroup
              }

\renewcommand{\subsection}
              [1]
              {
               \bgroup
               \flushleft
               \small\em
               \stepcounter{subsection}
               \arabic{section}.
               \arabic{subsection}. #1
               \par
               \egroup
              }

\renewcommand{\subsubsection}
              [1]
              {
               \bgroup
               \flushleft
               \small\em
               \stepcounter{subsubsection}
               \arabic{section}.
               \arabic{subsection}.
               \arabic{subsubsection}. #1
               \par
               \egroup
              }

  \newcommand{\acknowledgement}
              [1]
              {
               \bgroup
               \flushleft
               \small\bf
               #1
               \par
               \egroup
              }

  \newcommand{\sectionbib}
              [1]
              {
               \bgroup
               \flushleft
               \small\bf
               #1
               \par
               \egroup
              }

\newcommand{\lp}{\left(}
\newcommand{\rp}{\right)}

\newcommand{\pdt}[2][]{\partial_{#1}\lp{#2}\rp}

\newcommand{\pdn}[1]{\nabla\cdot\lp{#1}\rp}
\newcommand{\pdns}[1]{\nabla\lp{#1}\rp}

\newcommand{\reff}[1]{Fig.~\ref{#1}}

\newcommand{\refe}[1]{Eq.~\eqref{#1}}

\newcommand{\vect}[1]{\mathbf{#1}}

\newcommand{\avg}[1]{\langle{#1}\rangle}

\makeatletter
\DeclareFontEncoding{LS2}{}{\@noaccents}
\makeatother
\DeclareFontSubstitution{LS2}{stix}{m}{n}
\DeclareSymbolFont{largesymbolsstix}{LS2}{stixex}{m}{n}
\DeclareMathDelimiter{\lbrbrak}{\mathopen}{largesymbolsstix}{"EE}{largesymbolsstix}{"14}
\DeclareMathDelimiter{\rbrbrak}{\mathclose}{largesymbolsstix}{"EF}{largesymbolsstix}{"15}

\graphicspath{{./figures/}}
\renewcommand{\vec}[1]{\boldsymbol{#1}}
\renewcommand{\vect}[1]{\boldsymbol{#1}}

\newcommand{\rebuttal}[1]{{#1}}
\newcommand{\press}[1]{{#1}}
\newcommand{\picsize}{\small}
\newcommand{\picbox}[1]{{#1}}

\begin{document}

\title{\LARGE Applying Physics-Informed Enhanced Super-Resolution Generative Adversarial Networks to Turbulent Premixed Combustion and Engine-like Flame Kernel Direct Numerical Simulation Data}

\author{{\large Mathis Bode$^{a,b,*}$, Michael Gauding$^{b}$, Dominik Goeb$^{b}$, Tobias Falkenstein$^{b}$, Heinz Pitsch$^{b}$}\\[10pt]
        {\footnotesize \em $^a$J\"ulich Supercomputing Centre, Forschungszentrum J\"ulich GmbH, Germany}\\[-5pt]
        {\footnotesize \em $^b$Institute for Combustion Technology, RWTH Aachen University, Germany}}

\date{}


\small
\baselineskip 10pt


\twocolumn[\begin{@twocolumnfalse}
\vspace{50pt}
\maketitle
\vspace{40pt}
\rule{\textwidth}{0.5pt}
\begin{abstract} 
Models for finite-rate-chemistry in underresolved flows still \press{pose} one of the main challenges for predictive simulations of complex configurations. The problem gets even \press{more challenging} if turbulence \press{is} involved. This work advances the recently developed PIESRGAN modeling approach to turbulent premixed combustion.  For that, the physical information processed by the network and considered in the loss function are adjusted, the training process is smoothed, and especially effects \press{from} density changes are considered. The resulting model \press{provides} good results for a priori and a posteriori tests on direct numerical simulation data of a fully turbulent premixed flame kernel. \press{The limits} of the modeling approach are discussed. Finally, the model is \press{employed} to compute further realizations of the premixed flame kernel, which are analyzed with a scale-sensitive framework \press{regarding} their cycle-to-cycle variations. \rebuttal{The work shows that the data-driven PIESRGAN subfilter model \press{can} very accurately reproduce direct numerical simulation data on much coarser meshes, which is hardly possible with classical subfilter models, and enables studying statistical processes more efficiently due to the smaller computing cost.}
\end{abstract}
\vspace{10pt}
\parbox{1.0\textwidth}{\footnotesize {\em Keywords:} Generative Adversarial Network; Direct Numerical Simulation; Large-Eddy Simulation; Premixed Combustion; Engine}
\rule{\textwidth}{0.5pt}
\vspace{10pt}
\end{@twocolumnfalse}] 


\clearpage

\section{Introduction} \addvspace{10pt}
The availability of larger and larger, often extensively labeled datasets, either from large scale experiments, social media interactions, or simulations, has massively boosted the development of algorithms, software, and hardware to efficiently use data. For example, modern graphics processing units (GPUs) feature more than 100 TeraFLOPS FP16 performance, and state-of-the-art algorithms and implementations \press{can} almost fully employ this potential. As \press{a} consequence, data-driven methods have evolved as \press{an} important tool in many areas, enabling new use cases not possible a few years ago~\cite{wang2019,bhati2021,Vinyals2019}. They also emerged as useful tools, such as machine learning (ML) and deep learning (DL), in fluid dynamics and reactive flows applications, e.\,g., for replacing simpler algebraic closure models in Reynolds-averaged Navier-Stokes (RANS) simulations and large-eddy simulations (LESs) \cite{fukami2019synthetic} and for efficiently storing complex data \cite{ihme2009,bode2019bspline}. 

Technically, \press{the} ML and DL approaches rely on continuously updating network weights in a data-fed training process to minimize loss functions.  A particular combination of two DL networks, generator and discriminator, coupled by an adversarial loss term, is called \press{a} generative adversarial network (GAN)~\cite{goodfellow2014generative}. GANs aim to estimate the unknown probability density of observed data without an explicitly provided data likelihood function. These generative models implicitly learn by requiring access to \press{only} data samples from the unknown distribution. Therefore, the \press{GANs learning of} unknown data probability distribution is unsupervised and does not need any labels necessary in supervised learning scenarios. Finally, the trained generator network is used for prediction.

Bode et al.~\cite{bode2019,bode2020dev,bode2021dad,bode2022pdl} developed physics-informed enhanced super-resolution generative adversarial networks (PIESRGANs) by advancing ESRGANs (enhanced super-resolution GANs)~\cite{wang2018esrgan} to handle 3-D physical data and \press{add} physical information \press{in} the training process. \rebuttal{Furthermore, they performed an ablation study with turbulent data to find a suitable network architecture for physical problems.} The main idea of their model algorithm is to use high-fidelity data ("H"), e.\,g., from a fully resolved direct numerical simulation (DNS), and a corresponding filter operation to get pairs of high-fidelity data and filtered data ("F"). These pairs are used to train PIESRGAN to recover the high-fidelity data as accurately as possible based on the filtered data only. The generated data are called reconstructed data ("R"). Bode~\cite{bode2022dpl} recently extended this framework to laminar reactive flows with finite-rate-chemistry modeling by suggesting to solve additional transport equations on the reconstructed data and introducing an integrated lookup table approach to significantly reduce the time to \press{the} solution for parametric variations and the computational cost. They demonstrated the usability of this framework for a laminar lean premixed combustion case and received good results for emission prediction. However, Bode~\cite{bode2022dpl} did not use PIESRGAN on turbulent finite-rate-chemistry scalar fields. To fully focus on chemistry, they did not even \press{directly} train the network with the velocity fields.  More details including further relevant literature can be found in previous works~\cite{bode2021dad,bode2022dlc,bode2022dpl,bode2022spray}. The PIESRGAN \press{is extended} to turbulent finite-rate-chemistry data in this work for the first time \press{based} on the example of an engine-like premixed flame kernel.

Cycle-to-cycle variations (CCVs) in spark-ignition (SI) engines are still not fully understood, as many different in-cylinder processes are involved, interact complexly, and include a large range of spatial and temporal scales. However, CCVs are \press{known to be} affected by early flame kernel development~\cite{Zeng2019,Schiffmann2018}; faster early flame kernel growth correlates with faster combustion cycles. 3-D DNSs of early flame kernel development are possible and very well-suited to \press{derive} more insight into the complex phenomena involved~\cite{falkenstein2020dns,falkenstein2020le1, falkenstein2020le2,klein2006,uranakara2017}. One problem though are the costs of such simulations if they are run with finite-rate-chemistry and sufficiently high turbulence levels. As turbulence is a stochastic phenomenon, multiple realizations of flame kernels need to be run to accurately reproduce CCV behavior. Therefore, this is a very appealing application for complex data-driven model development such as PIESRGAN. \press{Particularly}, the reconstruction approach allows direct evaluation of the filtered density function (FDF) instead of relying on filtered-probability-functions, which are usually used in conventional models. Additionally, this \press{approach} enables the consideration of stochastic deviations of the FDF, which have not received much attention in the engine context yet. Accurate simulation data exist for training on one hand, but on the other, the cost per simulation needs \press{reduction} to \press{create} a statistically representative database for further analysis.

In this work, an iso-octane/air premixed flame kernel configuration, representing the underlying physical processes in real engines described above, is chosen as a first target case for the development of PIERSGAN for reactive turbulent finite-rate-chemistry flows. 

\section{Case setup of a turbulent premixed flame kernel} \addvspace{10pt}
Falkenstein et al.~\cite{falkenstein2020dns,falkenstein2020le1,falkenstein2020le2} computed a database of different iso-octane/air flame kernels, including fully turbulent kernels and planar cases\press{,} as well as cases with unity Lewis number and constant Lewis numbers.  They used a uniform mesh with 960 grid points per direction to discretize a periodic box, and a reaction mechanism with 26 species. The box was initialized with homogeneous isotropic turbulence (HIT), which decays over time and mimics the conditions in SI engines. The unburnt temperature is $T_\mathrm{u}=\SI{600}{\kelvin}$, the initial pressure $p^0=\SI{6}{\bar}$, \press{and} the air-fuel equivalence ratio $\phi=\num{1.0}$. Consequently, the \press{cases} \rebuttal{were designed to lie} in the thin reaction zone combustion regime. All were computed in the low-Mach limit using the Curtiss-Hirschfelder approximation~\cite{hirschfelder1964} for diffusive scalar transport and including the Soret effect.

To simplify the analysis, Falkenstein et al.~\cite{falkenstein2020dns} introduced a simplified reaction progress variable $\zeta$ with the thermal diffusion coefficient $D_\mathrm{th}$ as \press{the} diffusion coefficient reading
\begin{equation}
	\pdt[t]{\rho \zeta} + \pdn{\rho \vect{u} \zeta} = \pdn{\rho D_\mathrm{th} \pdns{\zeta}} + \dot{\omega}_\zeta
\label{dpt:eq:tran}
\end{equation}
with bold notation for vectors, $\nabla$ as del operator, and $\partial_t$ as time derivative. $\rho$ is the density, $\vect{u}$ the velocity vector, and $\dot{\omega}_\zeta$ the chemical source term of the simplified reaction progress variable, which is defined as
\begin{equation}
\dot{\omega}_\zeta = \dot{\omega}_{\mathrm{H}_2} + \dot{\omega}_{\mathrm{H}_2\mathrm{O}} + \dot{\omega}_{\mathrm{CO}} + \dot{\omega}_{\mathrm{CO}_2}.
\label{dpt:eq:source}
\end{equation}

\begin{figure}[!htb]
    \picsize
    \centering
        \begin{subfigure}[b]{20mm}
            \centering
            \includegraphics[width=\textwidth]{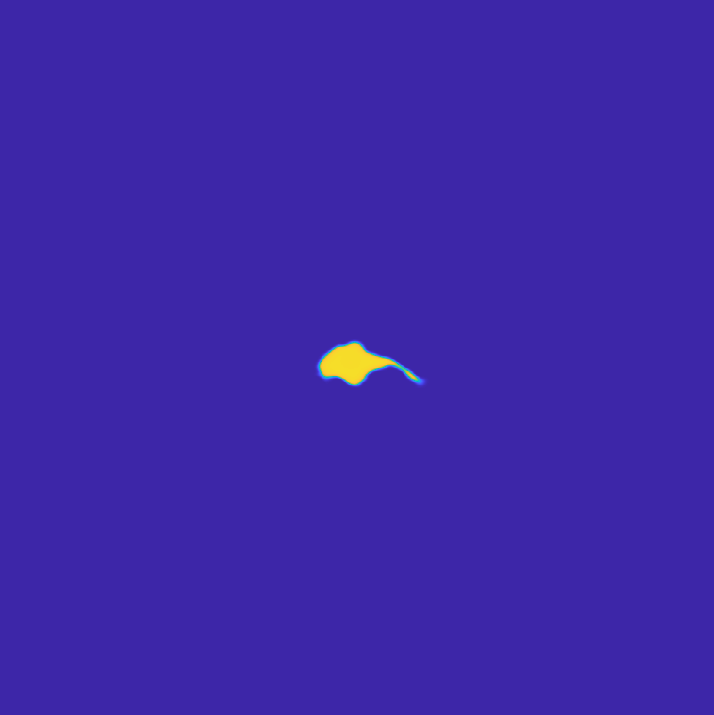}
            \label{dpt:sfig:t:11}
        \end{subfigure}
        \begin{subfigure}[b]{20mm}
            \centering
            \includegraphics[width=\textwidth]{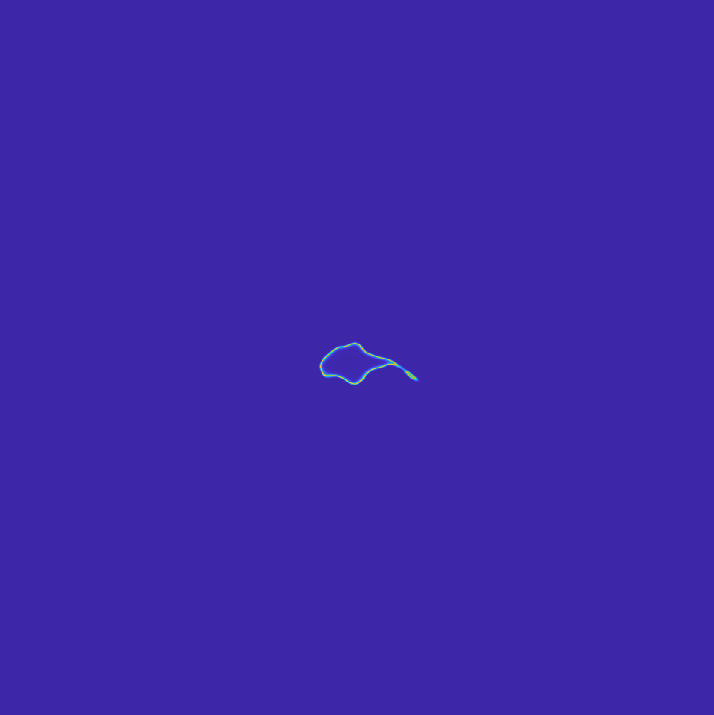}
            \label{dpt:sfig:t:12}
        \end{subfigure}
        \begin{subfigure}[b]{20mm}
            \centering
            \includegraphics[width=\textwidth]{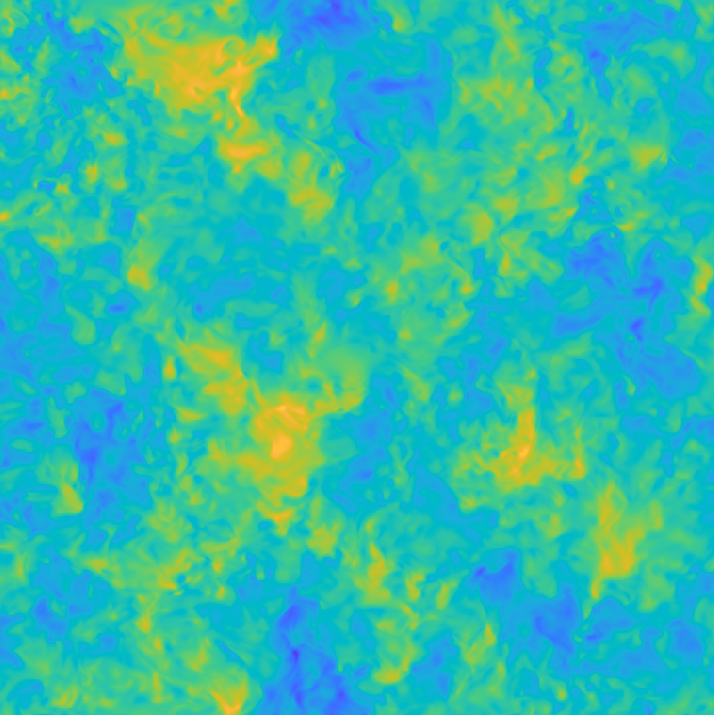}
            \label{dpt:sfig:t:13}
        \end{subfigure}
    \vskip -4mm
    
        \begin{subfigure}[b]{20mm}
            \centering
            \includegraphics[width=\textwidth]{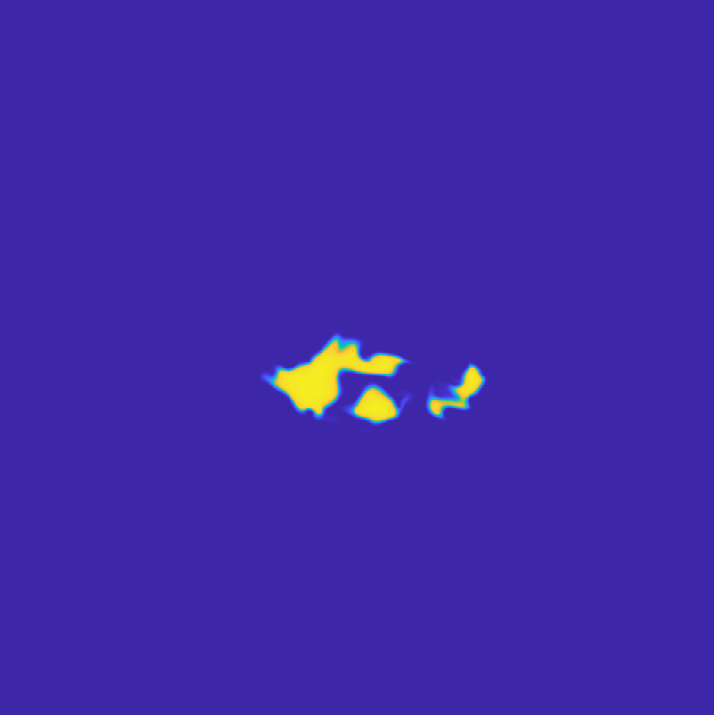}
            \label{dpt:sfig:t:21}
        \end{subfigure}
        \begin{subfigure}[b]{20mm}
            \centering
            \includegraphics[width=\textwidth]{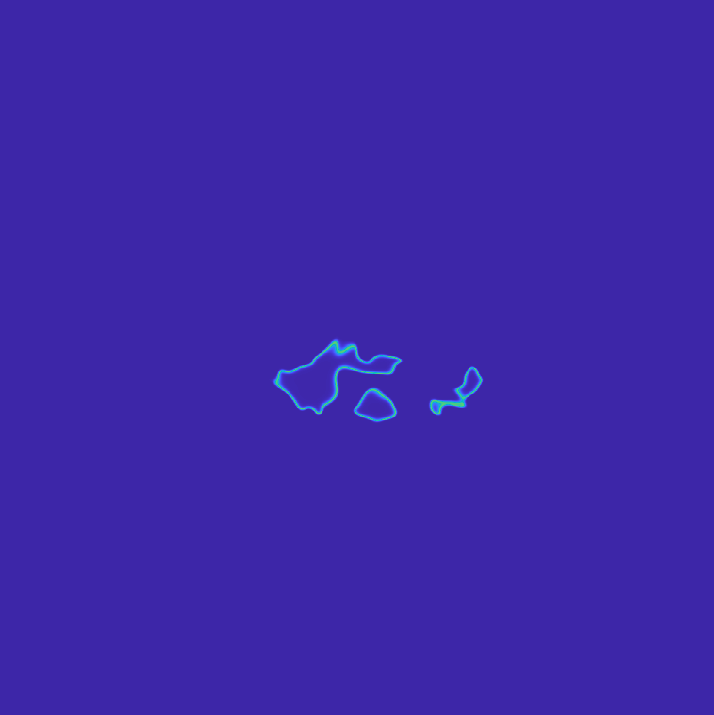}
            \label{dpt:sfig:t:22}
        \end{subfigure}
        \begin{subfigure}[b]{20mm}
            \centering
            \includegraphics[width=\textwidth]{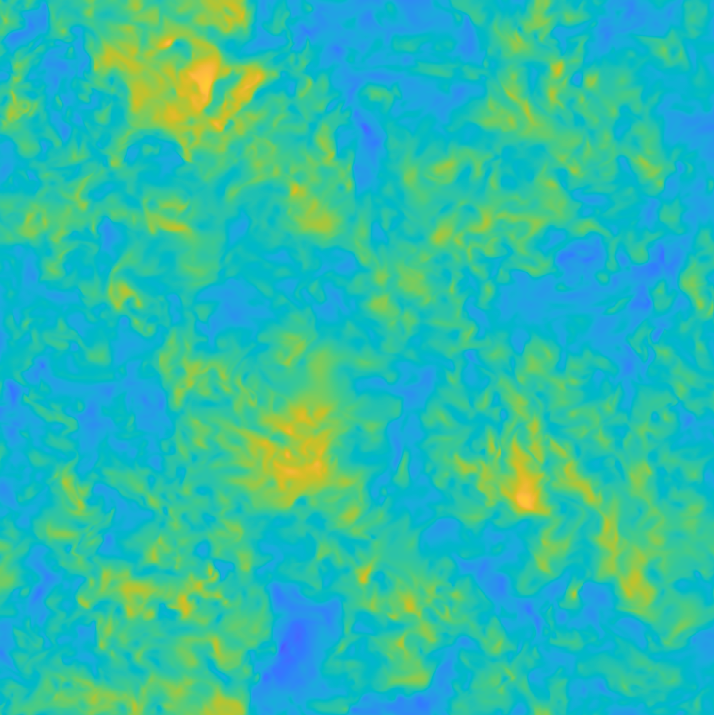}
            \label{dpt:sfig:t:23}
        \end{subfigure}
    \vskip -4mm
    
        \begin{subfigure}[b]{20mm}
            \centering
            \includegraphics[width=\textwidth]{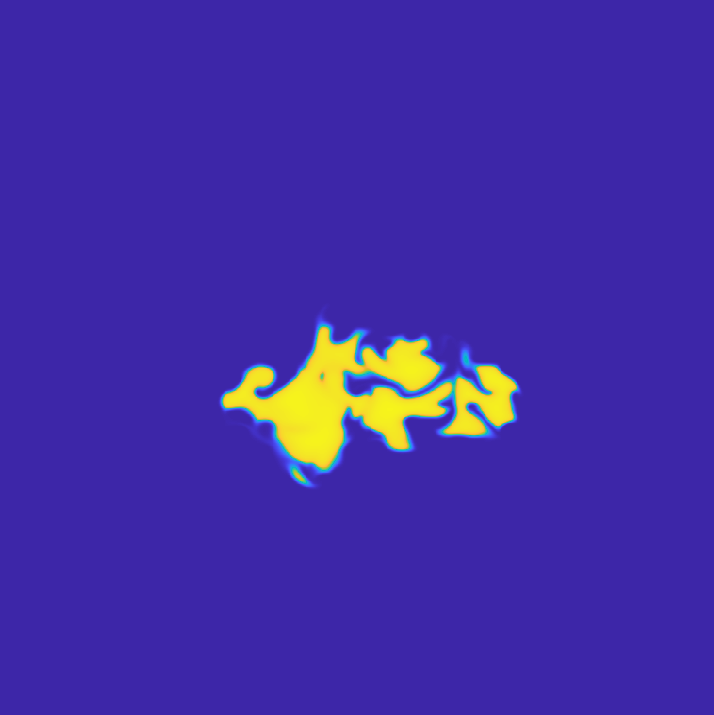}
            \label{dpt:sfig:t:31}
        \end{subfigure}
        \begin{subfigure}[b]{20mm}
            \centering
            \includegraphics[width=\textwidth]{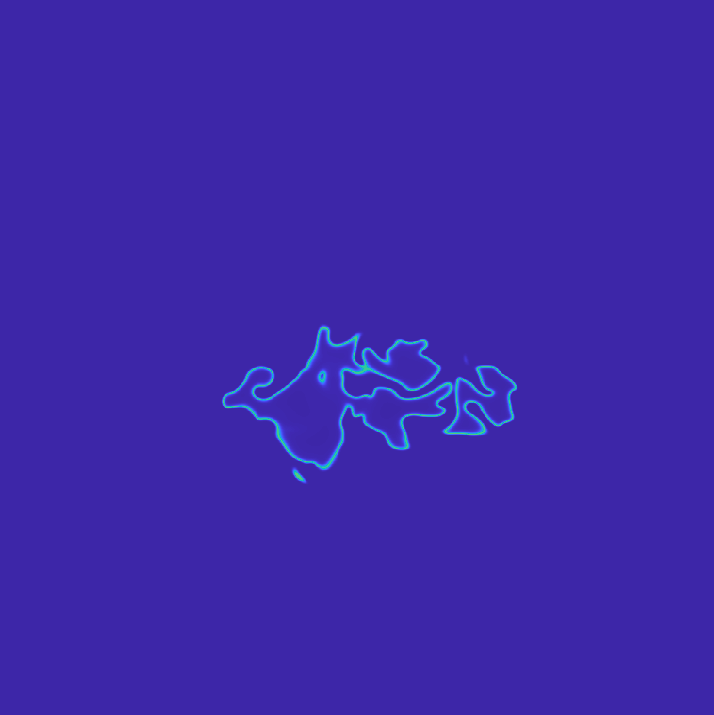}
            \label{dpt:sfig:t:32}
        \end{subfigure}
        \begin{subfigure}[b]{20mm}
            \centering
            \includegraphics[width=\textwidth]{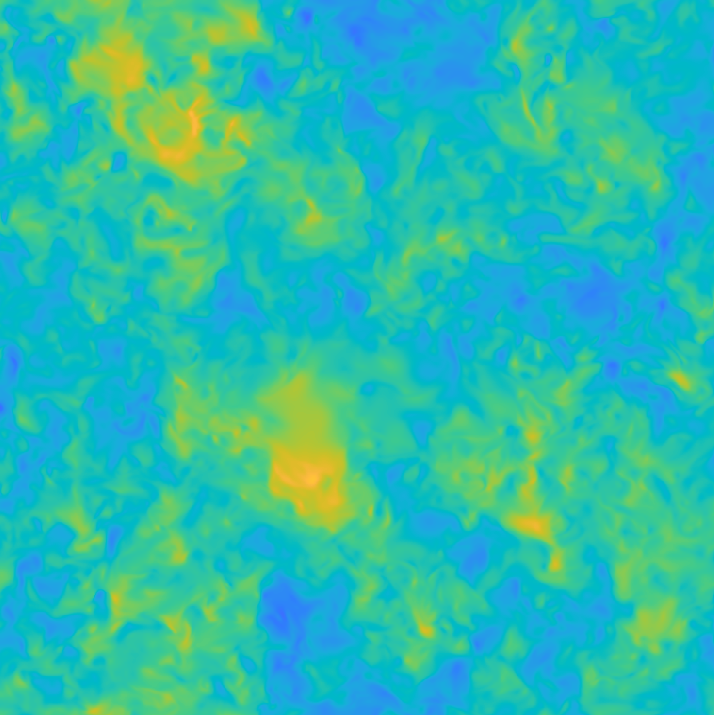}
            \label{dpt:sfig:t:33}
        \end{subfigure}
    \vskip -4mm
    
        \begin{subfigure}[b]{20mm}
            \centering
            \includegraphics[width=\textwidth]{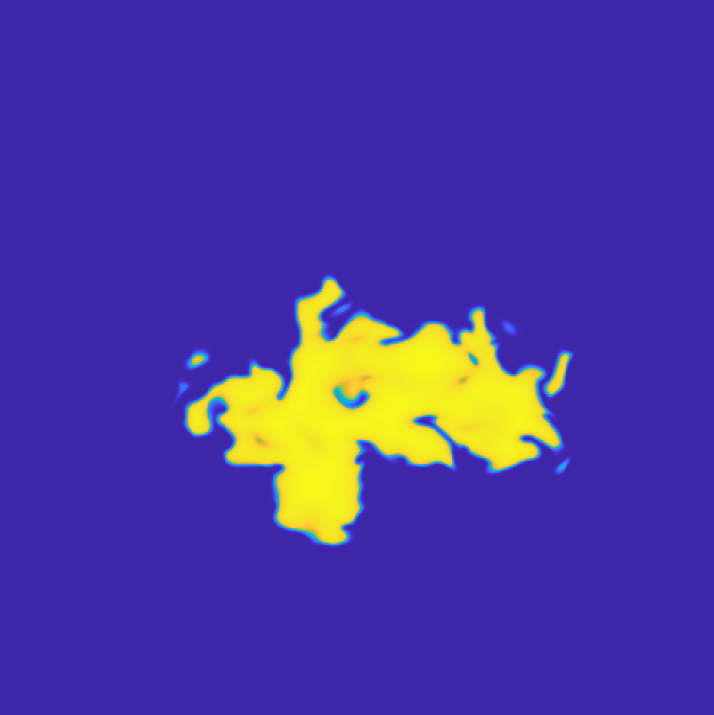}
            \label{dpt:sfig:t:41}
        \end{subfigure}
        \begin{subfigure}[b]{20mm}
            \centering
            \includegraphics[width=\textwidth]{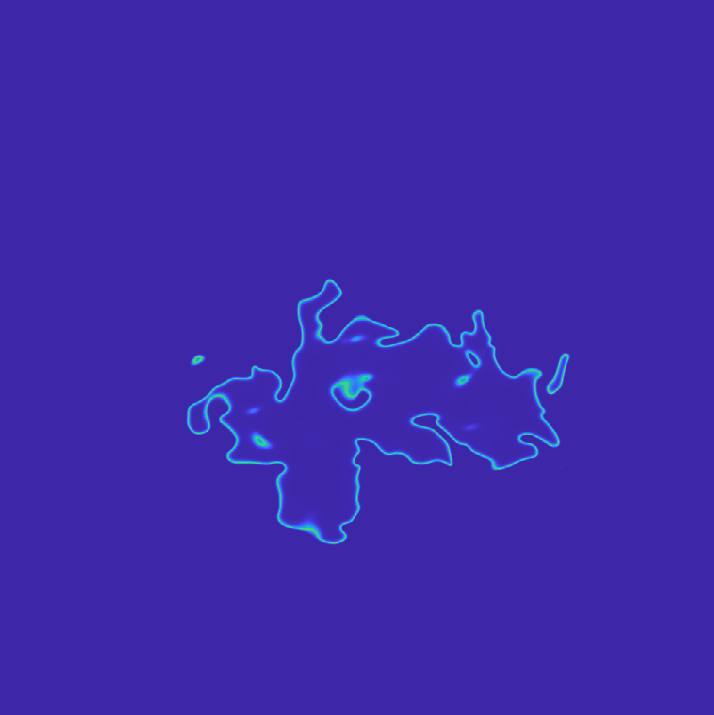}
            \label{dpt:sfig:t:42}
        \end{subfigure}
        \begin{subfigure}[b]{20mm}
            \centering
            \includegraphics[width=\textwidth]{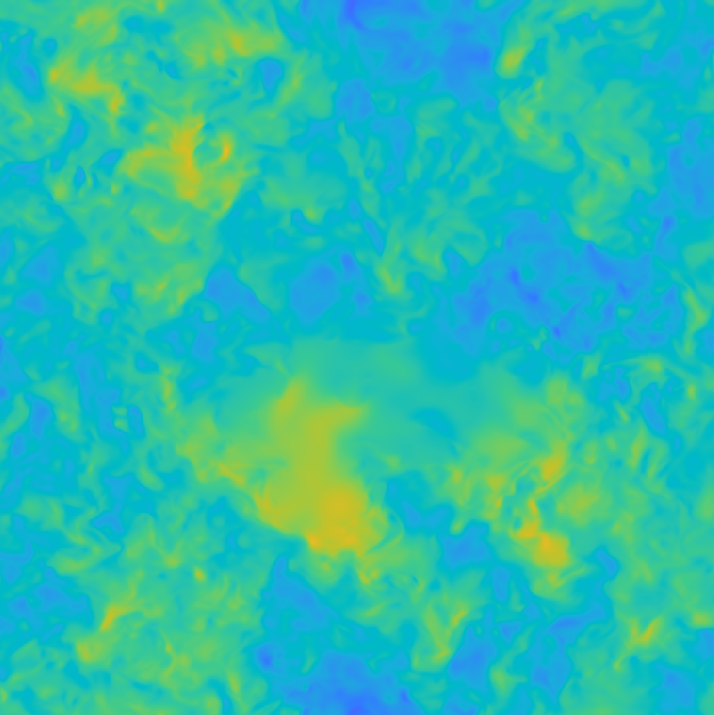}
            \label{dpt:sfig:t:43}
        \end{subfigure}
    \vskip -4mm
    
        \begin{subfigure}[b]{20mm}
            \centering
            \includegraphics[width=\textwidth]{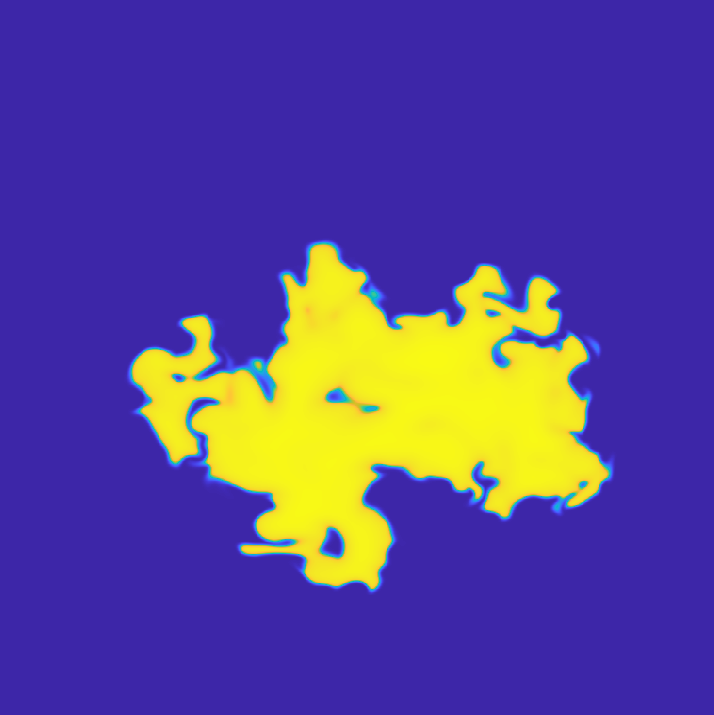}
            \label{dpt:sfig:t:51}
        \end{subfigure}
        \begin{subfigure}[b]{20mm}
            \centering
            \includegraphics[width=\textwidth]{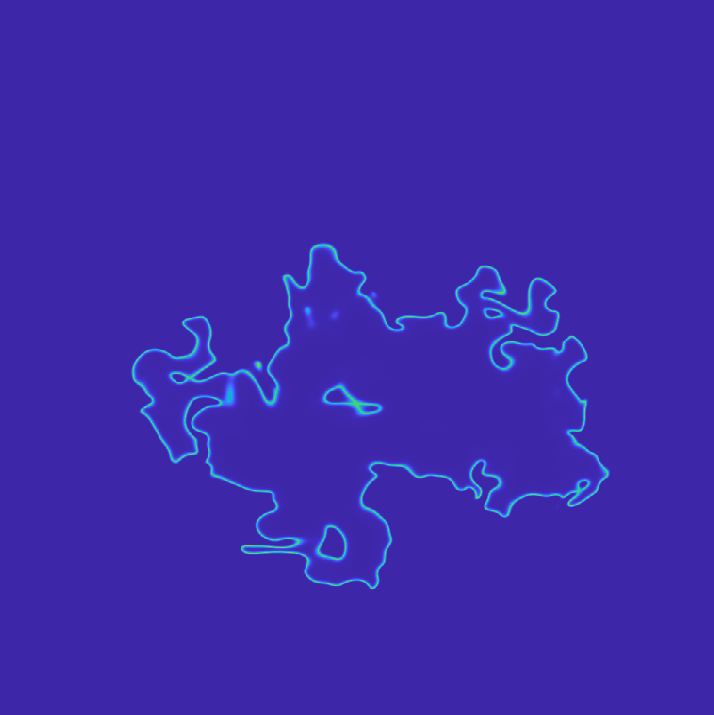}
            \label{dpt:sfig:t:52}
        \end{subfigure}
        \begin{subfigure}[b]{20mm}
            \centering
            \includegraphics[width=\textwidth]{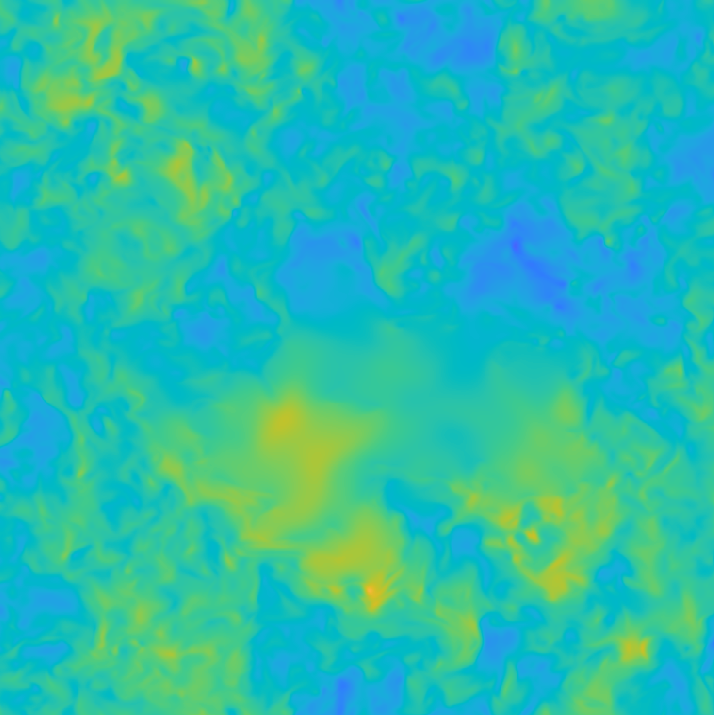}
            \label{dpt:sfig:t:53}
        \end{subfigure}
            \vskip -2 mm
       \caption{\rebuttal{Visualization of 2-D slices of $\zeta$, $\dot{\omega}_\zeta$, and $u_1$ (left to right) for five different increasing time steps (top to bottom: \SI{0.60E-4}{\second}, \SI{1.35E-4}{\second},  \SI{2.10E-4}{\second}, \SI{2.85E-4}{\second}, \SI{3.60E-4}{\second})} for the fully turbulent flame kernel with unity Lewis number. Note that the flame kernel does not break into parts at the latest time shown. A coherent flame kernel topology was maintained at all times.}
        \label{dpt:fig:vis}
    \end{figure}

To better understand the setup and its temporal development, 2-D slices through the center of the box are \press{presented} in \reff{dpt:fig:vis} for $\zeta$, $\dot{\omega}_\zeta$, and a velocity component $u_1$ for the case with unity Lewis numbers at three different time steps. As anticipated, over time, the kernel grows as shown by the larger area with higher $\zeta$-values and \press{wrinkles more}. Most of the reactions occur in the thin reaction zone. The turbulent structures indicated by the velocity component grow with time\press{,} implying a decrease of turbulent intensity in the box. Note that most of the analysis of this case in this work will be \press{performed} on later time steps than \press{are} presented here. These typically \press{offer} better statistical convergence due to a larger surface area.

\section{PIESRGAN for turbulent premixed combustion} \addvspace{10pt}
The PIESRGAN approach for finite rate chemistry by Bode~\cite{bode2022dpl}, except for the velocity and turbulence treatment\rebuttal{,} was mainly followed in this work. \rebuttal{The turbulent case considered here is much more challenging, as the chaotic nature of turbulence makes the training more difficult and the data much bigger, requiring more advanced computing approaches. Comparisons are \rebuttal{possible} \press{only} as ensemble statistics, which complicates the analysis.} It includes the reconstruction of fields, a consecutive solution step of the resolved transport equations on the finer mesh, the evaluation of the terms unclosed in the filtered equations on the coarser mesh, and the advancement of the filtered equations. \press{This algorithm is described in detail} in Bode~\cite{bode2022dpl}. \press{Moreover,} following their suggestion, \press{the} species \press{are split} into primary and secondary in this work. Transport equations are solved \press{only} for the primary species, while the secondary species are advanced by a chemtable-like lookup. One important reason is the reduced cost for the PIESRGAN with split species, which is thus called reduced PIESRGAN and denoted as PIESRGAN$_\mathrm{S}$. A PIESRGAN without secondary species is called \press{a} full PIESRGAN. \rebuttal{Details about the different aspects of PIESRGAN and \press{ways} to deal with the turbulence are \press{offered} in the next subsections.

\subsection{Loss function} \addvspace{10pt}
The loss function is the target function \press{that} is minimized by training the GAN. Bode~\cite{bode2022dpl} decoupled velocity and scalar fields \press{to train} their network for laminar partially premixed combustion. This \press{approach} is not possible here, as the fluctuations in the velocity field are expected to \press{significantly} impact all phenomena in the process. Thus, the loss function by Bode~\cite{bode2022dpl} needs to be complemented by a continuity loss term. However, this term cannot be similar to the \press{corresponding} term defined by Bode et al.~\cite{bode2021dad}\press{,} relying on a divergence-free condition for incompressible flows, which is obviously violated here. Consequently, the loss function for PIESRGAN for turbulent premixed combustion is chosen as
\begin{align}
\mathcal{L} = \beta_1 \press{L_\mathrm{adversarial}} + \beta_2 L_\mathrm{pixel} + \beta_3 L_\mathrm{gradient} \notag \\
+ \beta_4 L_\mathrm{continuity} + \beta_5 L_\mathrm{species},
\end{align}
where $\beta_1$, $\beta_2$, $\beta_3$, $\beta_4$, and $\beta_5$ are scalar coefficients weighting the different loss term contributions; in this work, these coefficients were always equally scaled such that the sum of all non-zero weighting coefficients \press{equalled} one. Note that all loss terms are non-dimensional, \press{as} all operators and input fields used are non-dimensionalized. The velocity field was zero mean-centered and rescaled with its root-mean-square deviation (RMSD) value. All loss terms are evaluated over all \press{the} cells of the domain. $L_\mathrm{adversarial}$ is the discriminator/generator relativistic adversarial loss~\cite{wang2019}\rebuttal{, which mainly communicates the discriminator feedback to the generator}. The pixel loss $L_\mathrm{pixel}$ and the gradient loss $L_\mathrm{gradient}$ are defined as the mean-squared error (MSE) of the quantity itself and of the gradient of the quantity, respectively~\cite{bode2019}. If the MSE operator is applied on tensors, including vectors such as the velocity, it is applied to all components separately. Afterward, the resulting tensor is mapped into a scalar using the $L_1$-norm. $L_\mathrm{continuity}$ is the physics-informed continuity loss, enforcing physically plausible solutions of the reconstructed flow field in which a compressible continuity equation should be fulfilled, $\pdt[t]{\rho} + \pdn{\rho \vect{u}}=0$. This \press{again contrasts} the original loss function by Bode et al.~\cite{bode2021dad} for HIT \press{in which} any dilatation and expansion effects were neglected. This introduces an additional complexity to the training process as two consecutive time steps were used to evaluate the temporal derivative. \press{To} increase the numerical accuracy, a centered derivative for time was employed, technically introducing an offset between \press{the} current filtered data and DNS data during the training process, as training has to wait for the availability of the next DNS data in time. $L_\mathrm{species}$ is the loss term \press{ensuring} that the sum of all mass fractions in the domain \press{equals} one.

The physics-informed loss terms, $L_\mathrm{continuity}$ and $L_\mathrm{species}$, were found essential for using PIESRGAN as \press{the} subfilter model. In the trained network, these terms must be very close to zero to prevent the \rebuttal{a posteriori} simulation \rebuttal{from blowing up}.

\subsection{Architecture} \addvspace{10pt}
The training process and architecture of PIESRGAN is shown in \reff{dpt:fig:piesrgan}. Both network parts, the generator and the discriminator, heavily utilize 3-D CNN layers (Conv3D) \cite{krizhevsky2012imagenet}, which are activated by leaky rectified linear unit (LeakyReLU) layers. The central component of the generator is the residual in residual dense block (RRDB). It contains residual dense blocks (RDBs) with skip-connections, which are extended residual blocks (RBs) with dense connections inside. A residual scaling factor $\beta_{\mathrm{RSF}}$ is used to prevent instabilities in the forward and backward propagation. The generator has about 80 layers in total. On the other hand, the discriminator has only about 28 layers. However, it features layers for batch normalization (BN) as well as dropout with dropout factor $\beta_{\mathrm{dropout}}$. Its final layer is a dense layer (Dense). The training is with pairs of data. The high-fidelity data ("H") are filtered in a prestep to receive filtered data ("F"), which \press{serve} as input to the network. The high-fidelity data are considered to evaluate the loss function terms.
\begin{figure*}[h!]
\picsize
\centering
\picbox{\includegraphics[width=\textwidth]{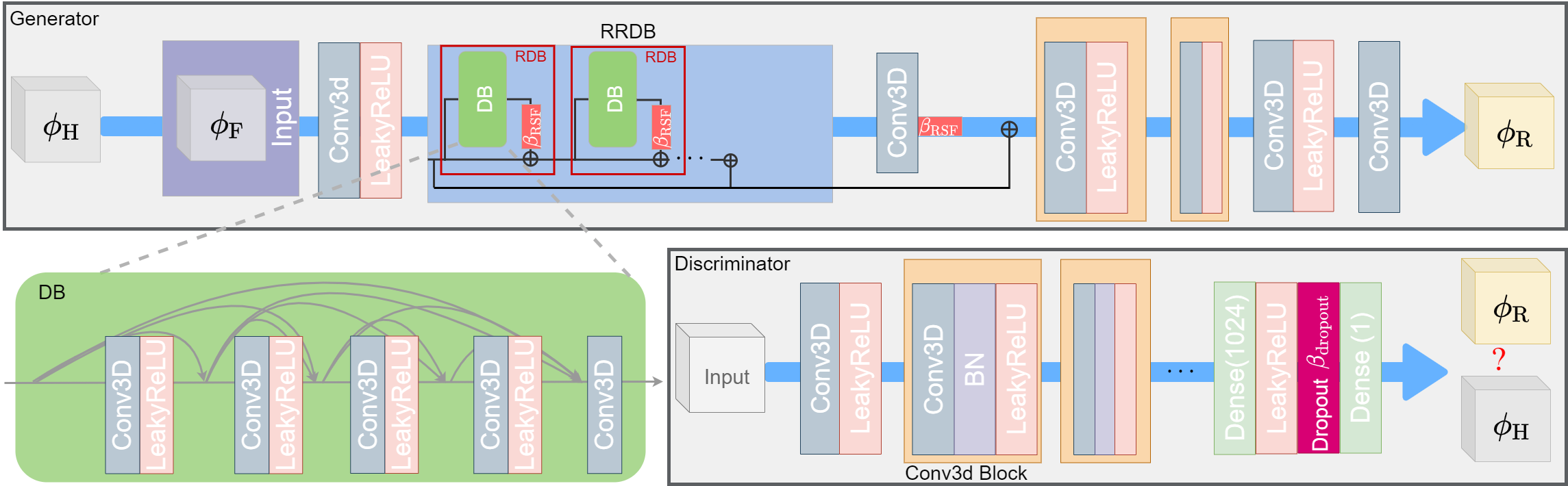}}
\caption{Sketch of PIESRGAN. "H" denotes high-fidelity data, such as DNS data, "F" are corresponding filtered data, and "R" are the reconstructed data.  The components are \press{as follows}: Conv3D - 3D Convolutional Layer, LeakyReLU - Activation Function, DB - Dense Block, RDB - Residual Dense Block, RRDB - Residual in Residual Dense Block, $\beta_\mathrm{RSF}$ - Residual Scaling Factor, BN - Batch Normalization, Dense - Fully Connected Layer, Dropout - Regularization Component, \press{and} $\beta_\mathrm{dropout}$ - Dropout Factor. Image from \cite{bode2021dad}.}
\label{dpt:fig:piesrgan}
\end{figure*}

\subsection{Species splitting} \addvspace{10pt}
For the turbulent premixed flame kernel application case in this work, a full PIESRGAN was trained first, which was then reduced by introducing primary and secondary species. Primary species are reconstructed on a fine mesh\press{,} and an unfiltered transport equation is solved to update the reconstructed field. The source term is evaluated on the fine mesh. On the other hand, secondary species are only reconstructed, including the source term. The accuracy is increased by shifting \press{the} primary and secondary species by a half\press{-}time step. The reduction \press{happened} using AutoML techniques \cite{hutter2019automated}. Simply put, species are systematically moved from primary to secondary, the effect is evaluated, and an optimal reduced PIESRGAN found.

In contrast to the findings by Bode~\cite{bode2022dpl}, $\mathrm{C}_8\mathrm{H}_{18}$, $\mathrm{OH}$, $\mathrm{H}_2$, $\mathrm{CO}$, $\mathrm{CO}_2$, and $\mathrm{H}_2\mathrm{O}$ were found to be the most suitable selection of primary species here. This is especially interesting for $\mathrm{H}_2$,  which was a secondary species for Bode~\cite{bode2022dpl}.

\subsection{Algorithm} \addvspace{10pt}
PIESRGAN is used to reconstruct "fully-resolved" data from filtered data. Therefore, the outputs of the GAN are refined data fields \press{used consecutively} to evaluate source terms and advance the filtered equations. The LES subfilter algorithm for the primary species starts with the LES solution without secondary species $\Phi_\mathrm{LES}^{n}$ at time step $n$ and consists of repeating the following steps:
\begin{enumerate}
\item Use the PIESRGAN to reconstruct $\Phi_\mathrm{R}^n$ from $\Phi_\mathrm{LES}^n$.
\item Use $\Phi_\mathrm{R}^n$ to update the primary species fields of $\Phi$ to $\Phi_\mathrm{R}^{n;\mathrm{update}}$ by evaluating the source terms and solving the unfiltered scalar equations on the mesh of $\Phi_\mathrm{R}^n$.
\item Use $\Phi_\mathrm{R}^{n;\mathrm{update}}$ to estimate the unclosed terms $\Psi_\mathrm{LES}^n$ in the LES equations of $\Phi$ by evaluating the local terms with $\Phi_\mathrm{R}^{n;\mathrm{update}}$ and applying a filter operator.
\item Use $\Psi_\mathrm{LES}^n$ and $\Phi_\mathrm{LES}^n$ to advance the LES equations of $\Phi$ to $\Phi_\mathrm{LES}^{n+1}$.
\end{enumerate}

For the secondary species and velocities, the second step is skipped. Any source terms are integrated in the network.

\subsection{Implementation and training details} \addvspace{10pt}
The training process is one of the main challenges for GANs. Depending on how it is done, e.\,g., how the network is initialized and how the learning rate is varied, the training might lead to a converging result or a diverging system. } In image applications, such as ESRGAN, a perceptual loss based on the VGG-feature space, which is pre-trained with, e.\,g., the ImageNet dataset, is often used to initialize the network coefficients and smoothen the training process.  A similar approach was followed in this work. The fully trained network by Bode et al.~\cite{bode2021dad} based on decaying HIT was used as \press{a} starting network to accelerate the training process.

The training was performed with multiple filter widths, using box filters for simplicity. The filter stencil width varied from 5 to 15 cells per direction. Furthermore, the training was performed on the fly to \press{efficiently use} compute nodes with GPUs, i.\,e., the Falkenstein et al.~\cite{falkenstein2020dns} configuration was rerun\press{,} and the obtained data was used for training without storing it permanently. This also minimized the probability \press{of} overfitting. \rebuttal{Moreover, this \press{always} allowed \press{comparison} between training and test data. Two newly run DNSs were used for on-the-fly training and the accuracy of the prediction compared to the original DNS data by Falkenstein et al.~\cite{falkenstein2020dns}.}

\rebuttal{The premixed flame kernel simulations were all run on a uniform mesh. Therefore, effects of non-uniform meshes were not further considered in this work. However, the presented method is not limited to uniform meshes. Bode~\cite{bode2022dnt} showed in the context of non-premixed flames that training with multiple filter widths can compensate mesh effects.}

The numerical solver and network parameters were chosen by Bode et al.~\cite{bode2021dad}, equally weighting $\beta_4$ and $\beta_5$.

\subsection{A priori testing} 
\begin{figure}[!htb]
    \picsize
    \centering
        \begin{subfigure}[b]{20mm}
            \centering
            \includegraphics[width=\textwidth]{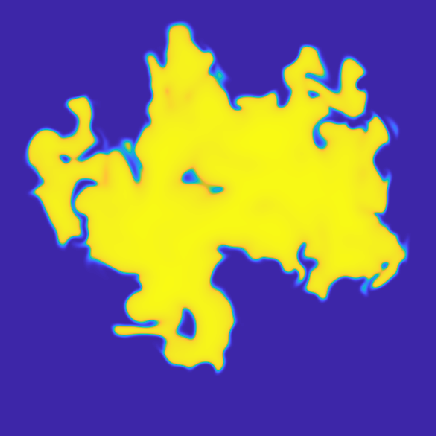}
            \vspace*{-10mm}
            \caption*{\colorbox{white}{$\zeta_\mathrm{H}$}}    
            \label{dpt:sfig:h:zeta}
        \end{subfigure}
        \hspace{-1mm}
        \begin{subfigure}[b]{20mm}
            \centering
            \includegraphics[width=\textwidth]{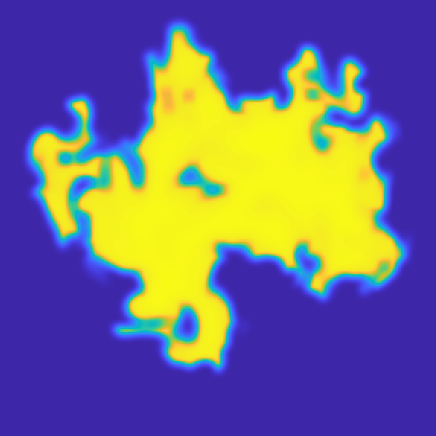}
            \vspace*{-10mm}
            \caption*{\colorbox{white}{$\zeta_\mathrm{F}$}}   
            \label{dpt:sfig:f:zeta}
        \end{subfigure}
        \hspace{-1mm}
        \begin{subfigure}[b]{20mm}
            \centering
            \includegraphics[width=\textwidth]{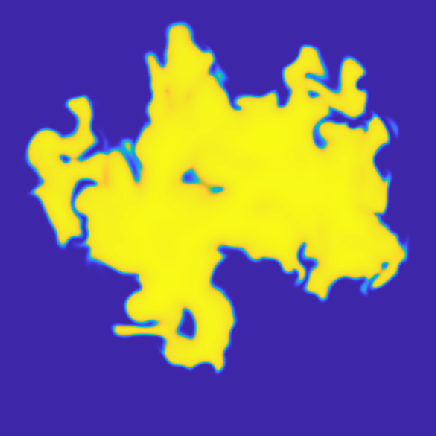}
            \vspace*{-10mm}
            \caption*{\colorbox{white}{$\zeta_\mathrm{RS}$}}   
            \label{dpt:sfig:rs:zeta}
        \end{subfigure}
    \vskip 1mm
    
        \begin{subfigure}[b]{20mm}
            \centering
            \includegraphics[width=\textwidth]{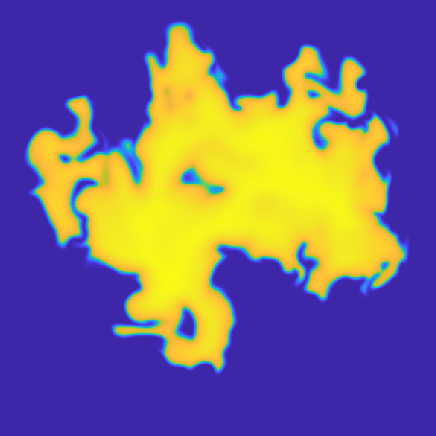}
            \vspace*{-10mm}
            \caption*{\colorbox{white}{$Y_\mathrm{C8H18;H}$}}    
            \label{dpt:sfig:h:C8H18}
        \end{subfigure}
        \hspace{-1mm}
        \begin{subfigure}[b]{20mm}
            \centering
            \includegraphics[width=\textwidth]{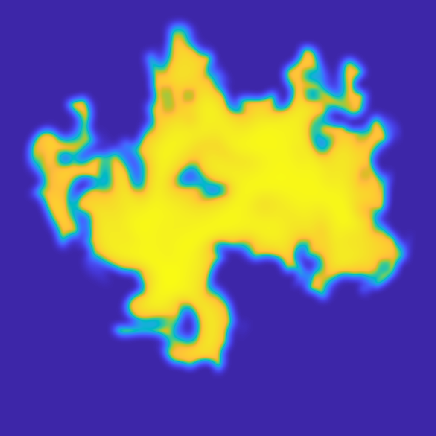}
            \vspace*{-10mm}
            \caption*{\colorbox{white}{$Y_\mathrm{C8H18;F}$}}   
            \label{dpt:sfig:f:C8H18}
        \end{subfigure}
        \hspace{-1mm}
        \begin{subfigure}[b]{20mm}
            \centering
            \includegraphics[width=\textwidth]{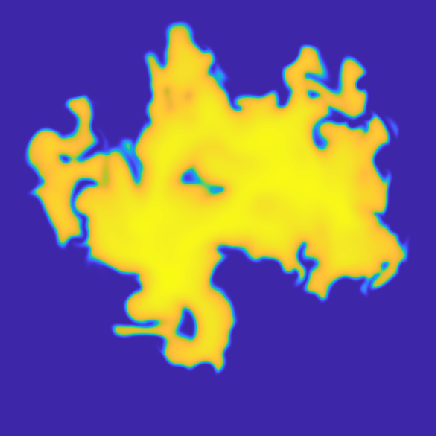}
            \vspace*{-10mm}
            \caption*{\colorbox{white}{$Y_\mathrm{C8H18;RS}$}}   
            \label{dpt:sfig:rc:C8H18}
        \end{subfigure}
    \vskip 1mm    
    
        \begin{subfigure}[b]{20mm}
            \centering
            \includegraphics[width=\textwidth]{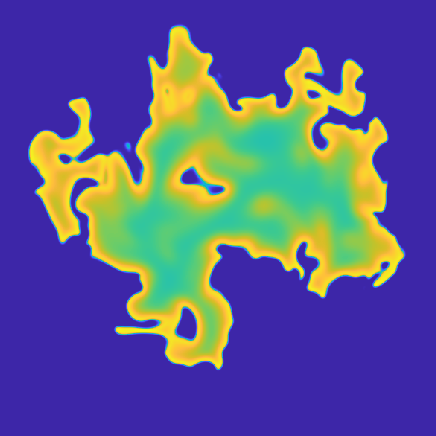}
            \vspace*{-10mm}
            \caption*{\colorbox{white}{$Y_\mathrm{OH;H}$}}    
            \label{dpt:sfig:h:OH}
        \end{subfigure}
        \hspace{-1mm}
        \begin{subfigure}[b]{20mm}
            \centering
            \includegraphics[width=\textwidth]{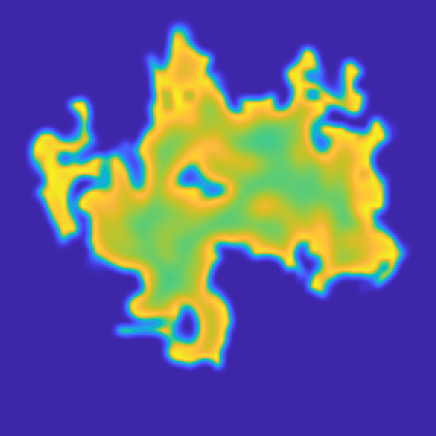}
            \vspace*{-10mm}
            \caption*{\colorbox{white}{$Y_\mathrm{OH;F}$}}   
            \label{dpt:sfig:f:OH}
        \end{subfigure}
        \hspace{-1mm}
        \begin{subfigure}[b]{20mm}
            \centering
            \includegraphics[width=\textwidth]{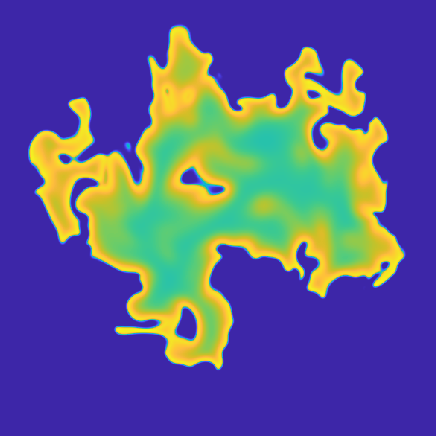}
            \vspace*{-10mm}
            \caption*{\colorbox{white}{$Y_\mathrm{OH;RS}$}}   
            \label{dpt:sfig:rs:OH}
        \end{subfigure}
    \vskip 1mm    

        \begin{subfigure}[b]{20mm}
            \centering
            \includegraphics[width=\textwidth]{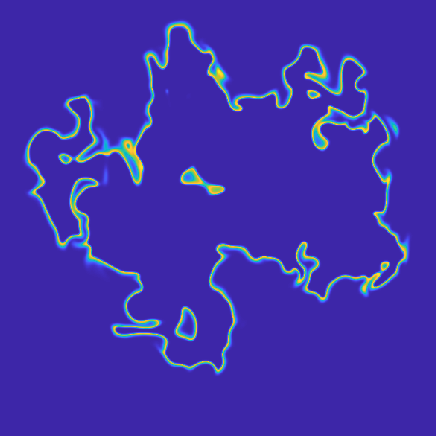}
            \vspace*{-10mm}
            \caption*{\colorbox{white}{$Y_\mathrm{CH2O;H}$}}    
            \label{dpt:sfig:h:CH2O}
        \end{subfigure}
        \hspace{-1mm}
        \begin{subfigure}[b]{20mm}
            \centering
            \includegraphics[width=\textwidth]{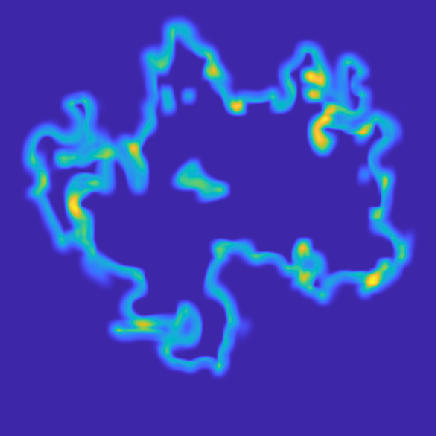}
            \vspace*{-10mm}
            \caption*{\colorbox{white}{$Y_\mathrm{CH2O;F}$}}   
            \label{dpt:sfig:f:CH2O}
        \end{subfigure}
        \hspace{-1mm}
        \begin{subfigure}[b]{20mm}
            \centering
            \includegraphics[width=\textwidth]{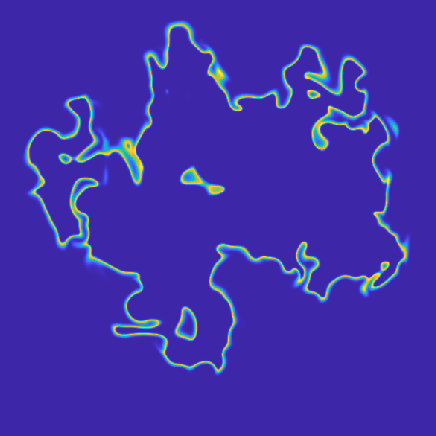}
            \vspace*{-10mm}
            \caption*{\colorbox{white}{$Y_\mathrm{CH2O;RS}$}}   
            \label{dpt:sfig:rs:CH2O}
        \end{subfigure}
    \vskip 1mm  
    
        \begin{subfigure}[b]{20mm}
            \centering
            \includegraphics[width=\textwidth]{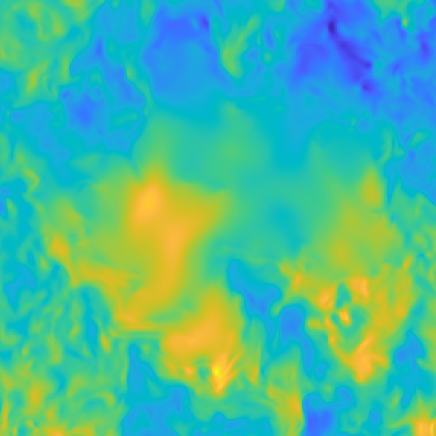}
            \vspace*{-10mm}
            \caption*{\colorbox{white}{$U_\mathrm{H}$}}    
            \label{dpt:sfig:h:U}
        \end{subfigure}
        \hspace{-1mm}
        \begin{subfigure}[b]{20mm}
            \centering
            \includegraphics[width=\textwidth]{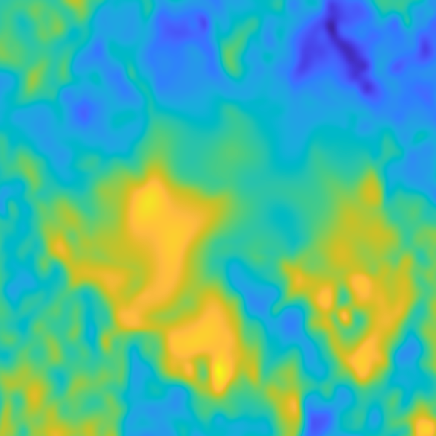}
            \vspace*{-10mm}
            \caption*{\colorbox{white}{$U_\mathrm{F}$}}    
            \label{dpt:sfig:f:U}
        \end{subfigure}
        \hspace{-1mm}
        \begin{subfigure}[b]{20mm}
            \centering
            \includegraphics[width=\textwidth]{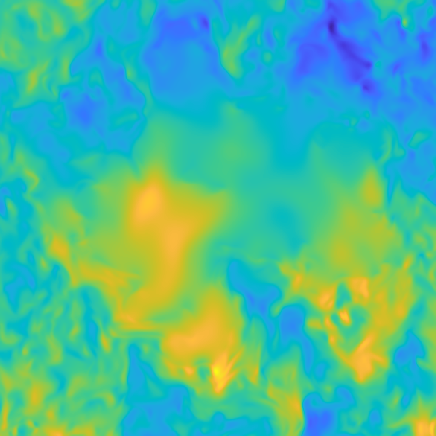}
            \vspace*{-10mm}
            \caption*{\colorbox{white}{$U_\mathrm{RS}$}}    
            \label{dpt:sfig:rs:U}
        \end{subfigure}
            \vskip 1 mm
       \caption{Visualization of DNS, filtered, and reconstructed fields for the unity Lewis number case employing PIESRGAN$_\mathrm{S}$. Results for the simplified reaction progress variable, the $\mathrm{C}_8\mathrm{H}_{18}$ mass fraction, the $\mathrm{OH}$ mass fraction, \rebuttal{the $\mathrm{CH}_2\mathrm{O}$ mass fraction}, and a velocity component are shown. Note that the images show the same time step as the last row in \reff{dpt:fig:vis} but are zoomed in. \rebuttal{Furthermore, all images show 2-D slices of the full 3-D data}.}
        \label{dpt:fig:recon}
    \end{figure}

\addvspace{10pt}
The accuracy of the $\mathrm{PIESRGAN}_\mathrm{S}$ model for turbulent premixed combustion super-resolution is first evaluated based on an a priori test, and the results are shown in \reff{dpt:fig:recon} for three species, a velocity component, and the simplified reaction progress variable. The reconstruction results of $\mathrm{PIESRGAN}_\mathrm{S}$ are denoted with "RS". The a priori test used the maximum filter stencil width of 15 cells per direction, while the ratio of cell size and Kolmogorov length on the fine mesh is about \num{0.85} at the considered time step. The agreement between \press{the} DNS data and \press{the} reconstructed data is very good. The filtered data are less sharp as they feature less small-scale structures, and it can be expected that running the simulation directly on such \press{a} coarse grid will result in unsatisfactory results. \rebuttal{Note that while $\mathrm{C}_8\mathrm{H}_{18}$ and $\mathrm{OH}$ are primary species, $\mathrm{CH}_2\mathrm{O}$ is a secondary species and also shows good accuracy.}

\subsection{A posteriori testing} 
\begin{figure}[!htb]
\picsize
	\centering
    \picbox{\input{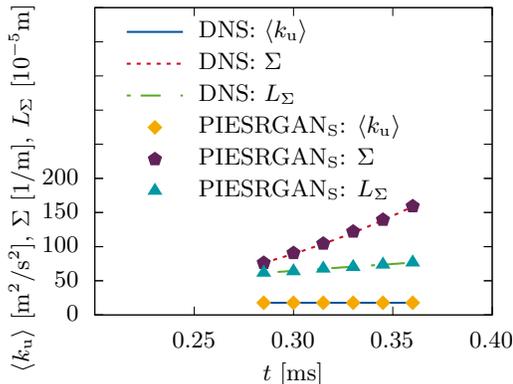}}
    \vskip 1mm
	\caption{Temporal evolution of the averaged turbulent kinetic energy in the unburnt mixture $\avg{k_\mathrm{u}}$, the surface density $\Sigma$, and the characteristic length scale $L_\Sigma$ for the DNS and PIESRGAN$_\mathrm{S}$ for the unity Lewis number case.}
	\label{dpt:fig:time}
\end{figure}

\addvspace{10pt}
To assess the accuracy of the trained PIESRGAN$_\mathrm{S}$ model, an a posteriori test has been conducted. For that, and to avoid different initial conditions, one of the already developed solutions of the DNS case with unity Lewis numbers, which was not used for training, was advanced in time by an LES employing the PIESRGAN$_\mathrm{S}$ model. A constant filter width of five cells per direction was chosen, it was maintained that all gradients were numerically sufficiently resolved in the transportation step on the coarse mesh, and multiple domain-averaged quantities (denoted by $\avg{\cdot}$) were evaluated. Note that even for this relatively weak filtering, simulations without model were not running stably. In \reff{dpt:fig:time}, the evolutions of the average turbulent kinetic energy in the unburnt mixture $\avg{k_\mathrm{u}}$, the surface density $\Sigma$, and the characteristic length scale $L_\Sigma$ (both defined in the application section) are shown. Satisfactory agreement between DNS and PIESRGAN$_\mathrm{S}$-LES can be observed.

It can be concluded that the PIERSGAN$_\mathrm{S}$ model \press{can} produce LES results \press{that} very closely mimic DNS results but on a much coarser grid. \press{Particularly}, only few DNSs, covering the parameter\press{-}bounding box, are required to train the initial network, and subsequent realizations can be generated using the much cheaper $\mathrm{PIESRGAN}_\mathrm{S}$-LES. 

\subsection{FDF analysis} 
\begin{figure}[!htb]
\picsize
	\centering
    \picbox{\input{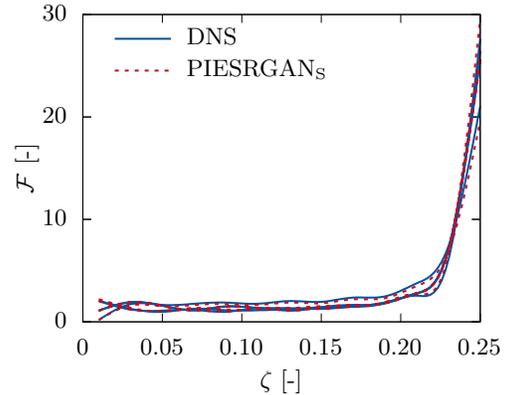}}
    \vskip 1mm
	\caption{FDFs \rebuttal{$\mathcal{F}$} for DNS data and PIESRGAN$_\mathrm{S}$ data in five different locations with the same filtered simplified reaction progress variable value.}
	\label{dpt:fdf:time}
\end{figure}

\addvspace{10pt}
In addition to the previously shown a posteriori test, another a posteriori test is performed. The target time step and model are the same, but a later start time is chosen\press{,} and the run time is only one fifth  of the previous a posteriori test. The filter width was increased to eleven cells per direction to enable statistically better computation of FDFs \rebuttal{(denoted $\mathcal{F}$)} with respect to the simplified reaction progress variable. FDFs are typically replaced by presumed PDFs in classical LES models. For example, a presumed PDF constructed by the filtered simplified reaction progress variable can be used to model the subfilter distribution, e.\,g., for evaluating the chemical source term. A consequence is that cells featuring the same value for the filtered simplified reaction progress variable (and potentially additional parameters) always have the same subfilter distribution, i.\,e., cannot account for stochastic deviations. In \reff{dpt:fdf:time}, FDFs for DNS data and PIESRGAN$_\mathrm{S}$ data in five different locations are compared. All selected locations feature the same filtered simplified reaction progress variable, which was evaluated as \press{the} progress variable value with maximum heat release. Deviations between the FDFs of different locations are obvious. This is interesting\press{,} and its impact on CCVs should be analyzed more systematically in the future. Another important result is that the agreement of the FDFs computed on the DNS data and the PIESRGAN$_\mathrm{S}$ data is good. This \press{result} again emphasizes the accuracy of the PIESRGAN$_\mathrm{S}$ model and \press{reveals} the advantages of the field reconstruction compared to simplified models based on presumed PDF closures. 

\subsection{Full versus reduced PIESRGAN} \addvspace{10pt}
The results so far obtained in this work but also the laminar results by Bode~\cite{bode2022dpl} \press{provide} the impression that the full and the reduced PIESRGAN \press{approaches} \press{offer} equally good results most of the time. This \press{result} is obviously not the case for all possible variations. For example, a reduced PIESRGAN trained with the unity Lewis numbers flame kernel case clearly overpredicts the evolution of the flame kernel case with constant, but non-unity Lewis numbers. Due to the learnt evolution of \press{the} secondary species on the reconstructed grid, the effect of non-unity Lewis numbers resulting in smaller source terms is not sufficiently considered. This \press{effect} can be seen even though a Lewis number effect is indirectly fed into the network \press{through} the primary species, for which the transport equations on the fine mesh include the Lewis number. The results for the full PIESRGAN model are much better, but still not perfect in this case. Overall, the PIERSGAN$_\mathrm{S}$ approach to utilize direct network lookup for secondary species seems to lead to very good \rebuttal{intra-case} capabilities of the network \rebuttal{and is computationally advantageous. However, \press{the approach} is more case-specific than a} full model solving transport equations for all species. 

\section{Application} \addvspace{10pt}
Overall, a priori and a posteriori results \press{provide} strong confidence that the developed PIESRGAN$_\mathrm{S}$-LES model is able to accurately and quickly compute turbulent flame kernel realizations, even on smaller computing clusters as long as GPUs are available. Therefore, in this section, six new independent realizations of early flame kernel development \rebuttal{with an 8-times lower resolution per direction than the original DNS} were computed, and their CCVs are analyzed in this section. It is not possible to \rebuttal{easily} predict the speed-up due to different clusters used \rebuttal{featuring CPUs and GPUs as well as} partly strongly varying compute cost per time step. However, all six PIESRGAN$_\mathrm{S}$-LESs together were slightly cheaper in terms of mixed FLOPS than one fully resolved DNS realization. \rebuttal{Generally, the speed-up increases for cases with even more species and is further discussed by Bode~\cite{bode2022sra}.} The six flame kernel realizations featuring different initial locations in the domain and therefore different turbulence interactions, are used in this section to gain more insight into early flame kernel development from a physical and modeling point of view. Obviously, six additional kernel realizations are not \press{sufficient} for a full statistical analysis to understand the role of early flame kernel development in CCVs. However, due to brevity and the modeling focus of this paper, this \press{number} should be sufficient to demonstrate the usage of the developed model in a scale-sensitive analysis framework. Therefore, the discussion is also only limited to a single time step, ($t=\SI{0.33}{\milli\second}$) even though the simulations were run over time. The analysis is performed on the reconstructed fields, featuring the same resolution as the original DNS.

\press{Assuming} the laminar flamelet concept, the geometrical
properties of a flame kernel are required to accurately predict its
macroscopic features, such as flame speed and heat release. A key
quantity is the flame surface density $\Sigma$\press{, which} determines the
reaction rate \cite{bray1985unified}.  Due to the interaction with the
turbulent flow, the flame surface exhibits a wide range of different
scales, extending from the integral scales, which contain information
about initial and boundary conditions, down to the Kolmogorov length
scale, which is the dissipative cut-off scale for which a
quasi-universal statistical theory exists~\cite{frisch1995turbulence}. 

Given the progress variable $\zeta$ that varies in space $\vect{x}$, a
threshold $\zeta_0$ can be used to define an interface that separates the regions
where $\zeta (\vect{x}) > \zeta_0$ from the regions where
$\zeta (\vect{x}) < \zeta_0$. Consequently, for a given threshold $\zeta_0$, a phase
indicator function $\Gamma(\vect{x},t)$ can be defined as
$\Gamma(\vect{x},t) = \mathcal H(\zeta(\vec{x},t) - \zeta_0)$, with
$\mathcal H$ being the Heaviside step function.  To proceed, the
structure function of the phase indicator function, i.\,e.,
\begin{equation}
  \label{eq:sf}
  \avg{(\delta\Gamma)^2}(\vect{r}, t) = \avg{(\Gamma(\vect{x} + \vec{r},t)  -\Gamma(\vect{x},t) )^2 }
\end{equation}
with $\vect{r}$ as spatial distance vector, is introduced and finally computed by taking an angular average to facilitate a scale sensitive analysis of the geometry
of the surface of the flame kernel.

The connection between the indicator structure function and the
morphology of the flame kernel surface can be demonstrated by taking
the small-scale and large-scale limits. Kirste and Parod~\cite{Kirste1962} proved that for
interfaces of class $\mathcal C^2$, the small-scale limit of \refe{eq:sf}
is given by
\begin{equation}
  \label{eq:small}
  \lim_{r\to 0} \avg{(\delta \Gamma)^2} = \frac{\Sigma r}{2} + \mathcal O(r^2).
\end{equation}
Here, the surface density is given by
\begin{equation}
  \Sigma =   \avg{| \nabla \Gamma |}
  \label{eq:6}
\end{equation}
and quantifies the area of the iso-scalar surface of the flame divided
by the volume $V$. Note that the computation of the surface density by \refe{eq:6} is computationally very efficient and does not require the tessellation of the iso-surface or the definition of a level set function. In the large-scale limit,
$\avg{(\delta \Gamma)^2}$ tends to
\begin{equation}
  \label{eq:large}
  \lim_{r\to\infty}  \avg{(\delta \Gamma)^2} = 2 \avg{\Gamma} ( 1 - \avg{\Gamma} ),
\end{equation}
and is hence related to the volume $\avg{\Gamma}$
enclosed by the iso-surface. If a sufficient scale separation exists,
\refe{eq:sf} also provides information about the fractal
dimension of the flame kernel surface \cite{elsas2018geometry}.  Furthermore, \press{expressing} the surface density by a characteristic length scale \press{is customary}
\begin{equation}
  \label{eq:Lsigma}
  L_\Sigma = \frac{4 \avg{\Gamma} ( 1 -\avg{\Gamma} )   }{\Sigma},
\end{equation}
which is related to the wrinkling scale
$L_\Sigma^* = (4\Sigma_{\rm max})^{-1}$ of premixed flames, which falls between the Taylor microscale and the Kolmogorov scale as recently demonstrated by Kulkarni et al.~\cite{kulkarni2021reynolds}.

\begin{figure}[!tb]
\picsize
	\centering
    \picbox{\input{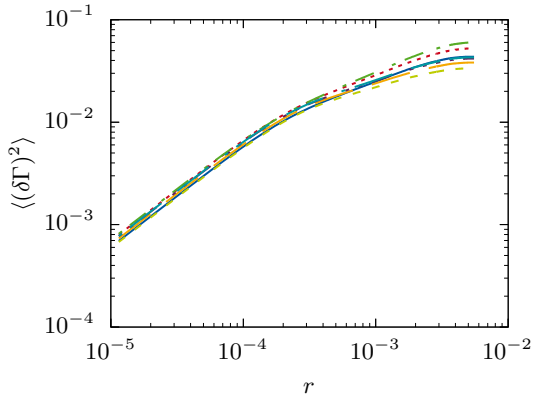}}
    \vskip 1mm
	\caption{Plot of the non-normalized phase indicator structure function for six realizations of the unity Lewis number case. Additionally, the non-normalized phase indicator structure function evaluated on the DNS data is shown (solid blue line).}
	\label{dpt:fig:sf}
\end{figure}

The discussed phase indicator structure function is an easily computable method to evaluate important multi-scale geometrical aspects of a flame kernel and to \press{understand} sensitivities for CCVs. For example, the surface density values evaluated from the six new realizations at the considered time step are
\SI{138.7}{\per\meter}, \SI{141.7}{\per\meter}, \SI{127.7}{\per\meter}, \SI{134.0}{\per\meter}, \SI{134.9}{\per\meter}, and \SI{117.7}{\per\meter}. The four largest values seem to be reasonably close together and the "standard" evolution, while the two smallest values deviate. Both deviating cases are slower in their evolution compared to the others, and thus, the smaller surface densities represent delayed flame kernels. This \press{fact} is also reflected in \reff{dpt:fig:sf}, which shows the non-normalized phase indicator structure function in log-log presentation. Additionally, \reff{dpt:fig:sf} shows the phase indicator structure function evaluated with the DNS data. It nicely aligns itself with the PIESRGAN$_\mathrm{S}$-LES computed realizations.

\begin{figure}[!tb]
\picsize
	\centering
    \picbox{\input{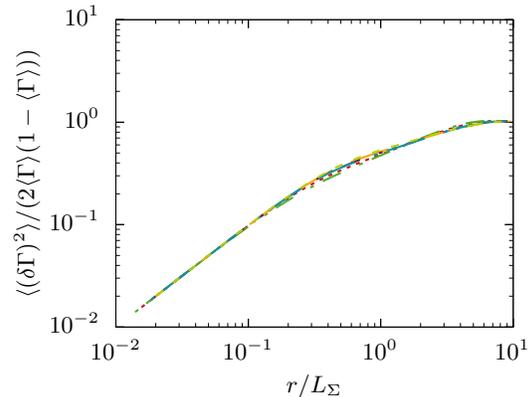}}
    \vskip 1mm
	\caption{Plot of the normalized phase indicator structure function for six realizations of the unity Lewis number case.}
	\label{dpt:fig:scaling}
\end{figure}

From a modeling perspective, \press{knowing} the characteristic scales of the phase indicator structure function leading to self-similarity \press{is important}. According to \refe{eq:Lsigma} and \refe{eq:large}, $L_\Sigma$ and $2 \avg{\Gamma} ( 1 - \avg{\Gamma} )$ are suitable candidates for scaling the phase indicator structure function. The result is \press{depicted} in \reff{dpt:fig:scaling}, which features excellent collapse at the small and large scales. At the intermediate scales, the influence of CCVs is clearly visible as variations of the flame kernel morphology.

\section{Conclusions} \addvspace{10pt}
This work shows the application of PIESRGAN-based modeling to turbulent finite-rate-chemistry flows for the first time.  The model for laminar finite-rate-chemistry flows described in Bode~\cite{bode2022dpl} is extended with special focus on the correct consideration of the fluctuating velocity fields and the density change, causing an expansion and deviating the turbulence from HIT. Both the full and the reduced PIESRGAN \press{models} \press{present} good results in a priori and a posteriori tests on \press{the} DNS data of a fully turbulent premixed flame kernel case under engine conditions. \press{The limits} of the reduced model were discussed. The differences between FDFs evaluated from the DNS or the reconstructed field from PIERSGAN and typical PDF closures were highlighted.

The data-driven model allows to compute more realizations of the fully turbulent premixed flame kernel for significantly lower cost. Six new realizations were computed for the unity Lewis number case for the cost of roughly one additional DNS realization and analyzed with respect to their CCV by means of a scale-sensitive framework. The level of variation was evaluated, and it was, furthermore, shown that the structure functions collapse sufficiently with appropriate scaling. This is an important conclusion from a modeling point of view as it helps develop simpler models connecting the flame surface density and heat release.

\press{Overall, this paper demonstrates that LES based on PIESRGAN subfilter modeling can very accurately predict complex turbulent finite-rate-chemistry flows. PIESRGAN probably outperforms all classical LES models.}

\acknowledgement{Acknowledgments} \addvspace{10pt}
\rebuttal{The authors acknowledge computing time grants for the projects JHPC55 and TurbulenceSL by the JARA-HPC Vergabegremium provided on the JARA-HPC Partition part of the supercomputer JURECA at J\"ulich Supercomputing Centre, Forschungszentrum J\"ulich,  the Gauss Centre for Supercomputing e.V. (www.gauss-centre.eu) for funding this project by providing computing time on the GCS Supercomputer JUWELS at Jülich Supercomputing Centre (JSC), and funding from the European Union's Horizon 2020 research and innovation program under the Center of Excellence in Combustion (CoEC) project, grant agreement no. 952181.}



 \footnotesize
 \baselineskip 9pt


\bibliographystyle{pci}
\bibliography{literature}


\newpage

\small
\baselineskip 10pt



\end{document}